# Spectral broadening in convex-concave multipass cells

Kilian Fritsch[1*], Victor Hariton[1,2,*], Kevin Schwarz[1], Nazar Kovalenko[1], Gonçalo Figueira[2] and Oleg Pronin[1]

[1]*Helmut-Schmidt-Universität / Universität der Bundeswehr Hamburg, Holstenhofweg 85, D-22043 Hamburg, Germany*
[2]*Instituto Superior Técnico, Universidade de Lisboa, Av. Rovisco Pais no1, 1049-001 Lisboa, Portugal*

*Corresponding author: kilian.fritsch@hsu-hh.de; victor.hariton@tecnico.ulisboa.pt

**Since its first demonstration in 2016, the multi-pass spectral broadening technique has covered impressive ranges of pulse energy (3 µJ – 100 mJ) and peak power (4 MW – 100 GW). Energy scaling of this technique into the joule-level is currently limited by phenomena such as optical damage, gas ionization and spatio-spectral beam inhomogeneity. These limitations can be overcome by the novel multi-pass convex-concave arrangement, which exhibits crucial properties such as large mode size and compactness. In a proof-of-principle experiment, 260 fs, 15 µJ and 200 µJ pulses are broadened and subsequently compressed to approximately 50 fs with 90 % efficiency and excellent spatio-spectral homogeneity across the beam profile.**

High energy, high-peak power ultrafast lasers are in demand for many applications such as laser-based plasma accelerators as well as particle, THz and X-ray sources [1,2]. In particular, high repetition rates and short pulse durations are sought after [1]. These are needed to make laser plasma accelerators compatible with real-world applications, and to improve existing scientific applications, for example, to push the cut-off wavelength of high-harmonic generation beyond the water-window and advance attosecond science. Industrial applications call for reasonable wall-plug efficiency and maintenance costs of their laser systems. This, in turn, implies the use of Yb-based, diode-pumped platforms. However, these laser systems are intrinsically limited in the shortest pulse duration achievable by the narrow bandwidth of the emission cross-section, requiring an external mechanism to shorten the pulses below the sub-ps level. Nonlinear spectral broadening via self-phase modulation (SPM) in hollow- or solid-core fibers followed by pulse compression is a well-established approach [3–5]. In 2016, the SPM-based spectral broadening was successfully transferred to free-space optics by Schulte et al. [6]. This geometry provides many new possibilities such as dispersion engineering, huge peak and average power adaptability, and scalability [7–11]. The technique is based on free-space propagation through a nonlinear medium (either dielectric bulk material or gases) in a multi-pass or quasi-waveguide arrangement [12]. The multi-pass setup repeatedly generates foci with sufficient intensity to accumulate SPM, while simultaneously distributing the nonlinear phase shifts over multiple passes with intermittent free-space propagation in the quasi-waveguide, which supresses the self-focusing and its detrimental effects. Dispersion compensation within the multi-pass arrangement as well as subsequent chirp removal of the SPM-induced phase leads to a temporal compression of the output pulses.

The highest energies at which this method has been successfully applied are 40 mJ, 5 kHz, 1.3 ps pulses compressed to 34 fs, and 112 mJ, 1.3 ps pulses spectrally broadened to an equivalent 37 fs Fourier-transform limit [9,13–15]. The applicability of the multi-pass method in the high-energy range is mainly limited by two detrimental effects: laser-induced damage of optical surfaces and ionization of the nonlinear material [16]. At high peak power levels (>$10^{12}$ W), ionization is currently the main bottleneck preventing further scaling [14]. So far, most of multi-pass cells implemented for the broadening experiments use a Herriott-cell (HC) arrangement with two concave mirrors (cav-cav configuration). The properties of their eigenmodes are determined by the radii of curvature (ROC) of each mirror and their separation. Increasing both parameters scales up the focus size with the square root of the cell length and thereby avoids ionization, although at the cost of an extremely large setup spanning up to ten meters [13,15].

HCs are similar in nature to optical resonators and are normally designed by using two concave mirrors. However, convex-concave (vex-cav) configurations can also be realized and are characterized by large mode volume [17]. This would allow an increase of the input energies toward values that were not achievable previously. No foci are present in the vex-cav multi-pass cell arrangement, which mitigates the ionization problem. In this configuration, virtual foci are located outside the cell behind the convex mirror M2 (see Figure 1). This has the twofold advantage of lowering the peak intensity and thus avoiding the ionization, while also allowing the setup to be folded multiple times, decreasing its length and making it manageable for practical applications. The folding is possible since the point of highest fluence is on the convex cell mirror, therefore a folding mirror may be inserted anywhere in the cell and experience lower fluences.

The vex-cav geometry was previously suggested for nonlinear multi-pass cells [16] with the emphasis on their spatio-temporal coupling. The experimental implementation was disregarded due to the lack of focus, keeping the beam in the far-field, and possibly leading to a spectrally inhomogeneous output beam profile. Here we experimentally demonstrate that the vex-cav configuration has excellent spectral homogeneity in the output beam, comparable to the well-established cav-cav arrangements. Following our patent application [18], this is the first demonstration of a vex-cav HC and its first use for applications in nonlinear optics, to the best of our knowledge. The approach is promising for achieving scalability in the joule energy range and making advancements in high-energy laser applications.

**A. Convex-Concave Setup**

The goal of this work is to provide an experimental comparison between the spectral broadening and compression results achieved with the vex-cav and with cav-cav HC geometry. Both HC schematics are depicted in Figure 1 and Table 1 outlines the main

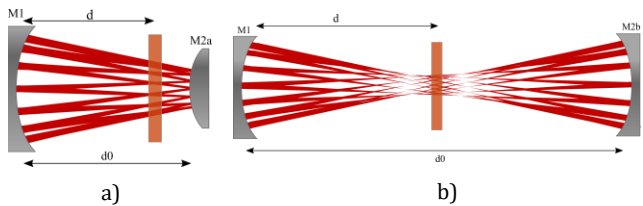

**Figure 1.** Schematic side view of a: a) vex-cav Herriott-type multi-pass cell and b) cav-cav multipass cell. Mirror M2a (shown here) corresponds to the convex mirror and M2b to the concave option, respectively, while M1 remains unchanged

parameters used and the results obtained in both configurations.
The vex-cav multipass cell was designed to broaden the output pulses from a commercial laser (Light Conversion PHAROS) delivering 15 W average power and 260 fs FWHM pulse duration at 1 MHz repetition frequency, corresponding to 56.4 MW peak power. A broadening factor of 5 was chosen due to a simple one-stage implementation and the possibility of comparison with the previously demonstrated configurations showing similar broadening factors [6–8]. The broadening setup consists of two HC mirrors and a fused silica plate for the nonlinear broadening medium. The left mirror (M1) is spherically concave with a -200 mm radius of curvature and 50.8 mm diameter. The right mirror (M2a) is spherically convex with a +250 mm ROC and 25.4 mm diameter. M1 has a dispersive coating introducing -140 fs$^2$ group delay dispersion (GDD) per reflection, in order to compensate the dispersion of the 3 mm thick fused silica plate. The coating on M2a is highly reflective (HR) with nearly zero dispersion. The separation between the two mirrors is adjusted to 114 mm, providing 19 reflections per mirror and 38 passes through the anti-reflective (AR) coated fused silica plate. The input and output coupling is performed by a rectangular scraper mirror in front of M1. The Gaussian eigenmode of the cell is defined by a beam radius of $w_1 = 336$ μm on M1 and $w_2 = 182$ μm on M2a. The laser mode is matched to the HC eigenmode by a Galilean-type beam expander. The fused silica plate is located at a distance d = 110 mm from M1. The eigenmode has a beam radius of w = 193 μm at this position (see Figure 1).

**Table 1.** Parameters used for spectral broadening in solid-state and gaseous nonlinear media.

| Multipass cell | Vex-Cav | Cav-Cav |
|---|---|---|
| Cell mirror ROCs (mm) | +250; -200 | -200; -250 |
| Distance between mirrors (mm) | 114 | 378 |
| Number of round trips | 38 | 38 |
| Spot waist (1/e$^2$) in focus (μm) | 193 | 195 |
| $P_{peak}/P_{crit}$ | ~14 | ~14 |
| $B_{int}$ per pass (rad) | ~0.51 | 0.5 |
| Fluence on mirror (mJ/cm$^2$) | 0.9; 2.9 | 0.76; 0.44 |
| Peak intensity on mirror (GW/cm$^2$) | 31; 103 | 27.5; 16 |
| Thickness FS (mm) | 3 | 3 |
| $n_2$ (m$^2$/W) | 2.3·10$^{-20}$ | 2.3·10$^{-20}$ |
| Input duration (fs) | 260 | 260 |
| Pulse energy (μJ) | 15 | 15 |
| Peak intensity in focus (TW/cm$^2$) | 0.10 | 0.13 |
| Output FTL duration (fs) | 49 | 53 |

The broadened output spectrum spans from 990 nm to 1070 nm (Figure 2a, shaded area) with a corresponding Fourier-transform limit (FTL) of 49 fs, which amounts to a compression factor of 5.5. Pulse compression is achieved by six dispersive mirrors with -400 fs$^2$ GDD each, resulting in a measured pulse duration of 53 fs FWHM (Figure 2b, solid line), with a transmission efficiency through the HC and the compression stage of 91%. Over 80% of the remaining energy is located in the main pulse, as shown by the retrieved trace of a frequency-resolved optical gating (FROG) device (pulseCheck 15, APE GmbH).

In order cross-check the experimental results with the vex-cav geometry, numerical simulations of nonlinear pulse propagation are conducted with the SISYFOS [19] code, which includes both transverse dimensions and the Kerr effect presented in the FS plate. The spatial grid was restricted to a single quadrant by exploiting the mirror symmetry of the beam. The code has been previously benchmarked and was applied to bulk and multi-pass cell (MPC) broadening simulations [20–22] in different energy ranges and configurations. The simulation results are shown in Figure 2; it uses a nonlinear refractive ($n_2$) index of 2.3×10$^{-20}$ m$^2$/W for the FS plate and assumes vacuum between the cell mirrors, neglecting the air nonlinearity. The code predicts accurately the amount of broadening and the output duration from this configuration.

In the current setup, a nonlinear phase of $\varphi \approx 0.25$ rad per pass is adopted while the gas-filled multi-pass cells can tolerate up to 1 rad per pass [14]. The measurement of the spatio-spectral homogeneity is particularly important in order to compare the performance of the vex-cav configuration with the cav-cav configuration. Experimentally this is assessed by comparing spectra of small samples of the beam profile selected by scanning a fiber tip over two orthogonal axes (more details in the Supplementary). We obtain $V_x^{avg} = 99\%$ and $V_y^{avg} = 98\%$ for the tangential and sagittal directions respectively, which are comparable to state-of-the-art cav-cav systems with similar broadening factors in bulk material. Moreover, the spectral overlap exceeds 90% in the beam diameter of ~3 mm defined as 1/e$^2$, which indicates a remarkable homogeneity. Similar values were found using the SISYFOS code, which predicts a homogeneity well above 90% (see Figure S1 in Supplementary material).

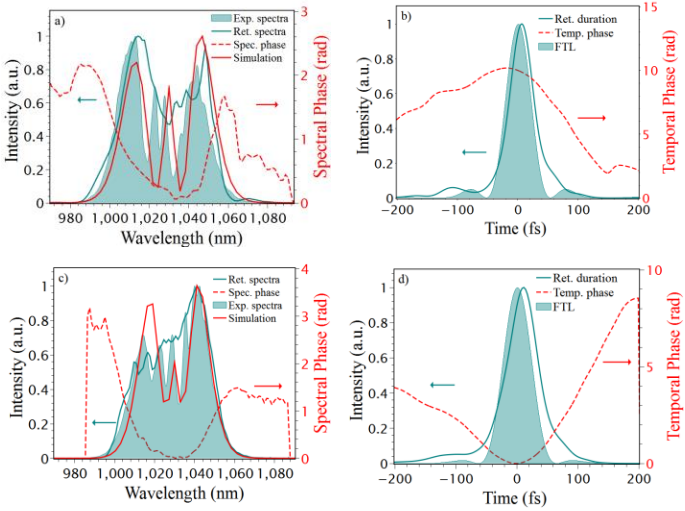

**Figure 2.** Measured (shaded area), simulated (red line) and FROG-retrieved (green line) output spectrum and spectral phase (dashed line) for a) the vex-cav configuration and c) cav-cav configuration, at an input energy of 15 µJ. b) Measured duration of 53 fs with a FTL of 49 fs for vex-cav cell. d) Measured duration of 57 fs with a FTL of 53 fs for cav-cav cell

### B. Concave-Concave setup

Even though the spectral homogeneity of the vex-cav system is comparable to those reported in other publications [8,13,23], we also set up a cav-cav configuration to make a direct comparison to this well-established geometry. The design goal was to find a cav-cav HC configuration providing the same broadening factor and enabling a compressed pulse duration of approximately 53 fs, comparable to the result obtained with the vex-cav geometry. Parameters such as the number of passes through the nonlinear material, the input pulse duration and energy, and the dispersion properties within the HC were preserved. The cell length was adjusted accordingly to yield a similar eigenmode. Mirror M1 was the same in both experiments, but for the cav-cav configuration we replaced mirror M2a by mirror M2b with -250 mm ROC and a highly reflective dielectric coating. The overall dispersion within the HC stayed approximately the same. The mirror separation was adjusted to 378 mm, which enabled the same 19 reflections per mirror and 38 passes through the nonlinear medium. The fused silica plate is placed at a distance d = 110 mm from M1. The Gaussian eigenmode of the cell is characterized by a beam radius of $w_1 = 358$ µm and $w_2 = 471$ µm on M1 and M2b respectively as well as w = 195 µm in the nonlinear material.

Figure 2c) shows the output beam spectrum. The spectrum spans from 990 nm to 1070 nm, with a corresponding FTL of 53 fs. Using the same dispersive mirror compressor with a total GDD compensation of -2400 fs$^2$, we achieve a compressed pulse duration of 57 fs FWHM, representing a pulse-shortening factor of 5. Figure 2d) shows the FROG retrieved temporal pulse profile. A 90% overall transmittance through the HC was measured. The spectrum is once again predicted using the simulation code. The calculated spectral overlap for both axes and the corresponding prediction from the simulations is plotted in Figure 3b). The calculated effective overlaps for this configuration are $V_x^{avg} = 99\%$ and $V_y^{avg} = 99\%$. These values are comparable to state-of-the-art measurements for spectral broadening in multipass systems. The V parameter values well below 90% could be measured in our lab when the cell was operated with a B-integral over 0.6 per pass.

Based on these measurements, the vex-cav configuration performs similarly to traditional cav-cav setups under the presented conditions. In either situation, the 260 fs input pulses were compressed to ~50 fs with similar efficiencies (over 90 %), which is in line with the design target and the data reported in the other publications. More remarkably, the overall homogeneity values $V_x^{avg}$ and $V_y^{avg}$ are both similar and above 98 % and it is accurately predicted by the simulations. The proposed vex-cav broadening cell combines the excellent stability and robustness of HC-based systems while still maintaining the beam homogeneity.

### C. Energy scaling in convex-concave multipass cells

Essentially, solid-state spectral broadening in cav-cav geometry is not limited by the input peak power and importantly by $P_{peak}/P_{crit}$. For example, previous demonstrations [6,7] showed $P_{peak}/P_{crit}$ to be in the range 1-20 and even up to 1800 [14]. Importantly, the B-integral per pass should stay in the range of <0.6 rad to avoid any spatial beam deteriorations. Here we experimentally demonstrate that the same considerations apply to vex-cav configurations and show its operation in the regime $P_{peak}/P_{crit} = 200$. The setup consists of a concave mirror with a ROC of -1000 mm and a convex mirror of +3000 mm ROC both with 25.4 mm diameter, which allows an increased mode size inside the cell. The mirrors have a HR coating with no GDD compensation. The separation between them is only L = 283 mm yielding N = 13 bounces per mirror. The Gaussian eigenmode of the cell is characterized by a beam radius of $w_1 = 497$ µm on the concave mirror and $w_2 = 402$ µm on the convex mirror. Importantly, compared to cav-cav cells the eigenmode of vex-cav cell has nearly no divergence. The maximum input energy into the cell was limited to 200 µJ out of a Light Conversion PHAROS laser. To confirm the feasibility, the 260 fs long pulses were broadened in two separated nonlinear media. First, a 1 mm FS plate placed 40 mm from the convex mirror and second, a pressurized environment with 7 bar of Argon. The input peak power is 720 MW resulting in a $P_{peak}/P_{crit}$ ratio of 200 in case of the FS (~3.6 MW) and a ratio of 0.33 for the Argon. The compilation of the used parameters is presented in the Table 2.

**Table 2.** Experimental parameters for a Herriott-cell configuration for an input energy of 200 µJ. (Cell configuration) $R_i$: ROC of cavity mirror i = **1**, **2**; $L_{cell}$: total cell length; $M_{rt}$: cell configuration number; $N_r$ : number of roundtrips. (eigenmode) ; η: efficiency; $w_i$ : beam width on cavity mirror i = **1**, **2**; $w_{egn}$: eigenmode waist in the focus; $B_{int}$: B-integral per pass; $I_{peak}$, $F_{peak}$ : peak intensity and fluence. (input pulse) $E_p$, $P_{peak}$, $\tau_{fwhm}$, $\lambda_0$, $P_{peak}/P_{crit}$: input pulse energy, peak power, duration, central wavelength and peak over critical power ratio.

| Cell configuration | | Eigenmode | | Input pulse | |
|---|---|---|---|---|---|
| $R_1$ | -1000 mm | $w_1$ | 497 µm | $E_p$ | 200 µJ |
| $R_2$ | +3000 mm | $w_2$ | 402 µm | $P_{peak}$ | 7.2 ×10$^8$ W |
| $L_{cell}$ | 283 mm | $w_{egn}$ | 397 µm | $\tau_{fwhm}$ | 260 ×10$^{-15}$ s |
| $M_{rt}$ | 2 | $F_{peak}$ | 7.9 ×10$^{-2}$ J cm$^{-2}$ | $\lambda_0$ | 1030 nm |
| $N_r$ | 13 | $I_{peak}$ | 2.6 ×10$^{11}$ W cm$^{-2}$ | $P_{peak}/P_{crit}$ (FS) | 200 |
| η [%] | 93 | $B_{int}$ | ~0.5 | $P_{peak}/P_{crit}$ (Ar) | 0.33 |

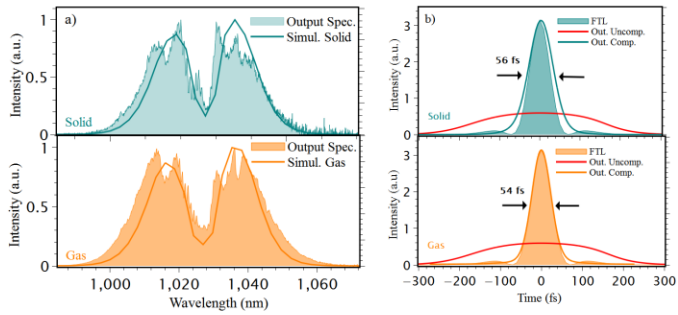

**Figure 3**. a) Measured (shaded area) and simulated (solid line) output spectra for the vex-cav configuration at an input energy of 200 µJ, for solid- (top) and gas-based (bottom) broadening. b) Measured output duration and the respective FTL comparison for solid and gas media.

A B-integral per pass of approximately 0.5 rad was possible, in both cases, leading to a spectral coverage from 980 to 1070 nm as shown in Figure 3a). An excellent homogeneity is preserved with both values $V_x^{avg}$ and $V_y^{avg}$ above 90 % (Figure 4), in the $1/e^2$ region, with the transmission over 93 %. Once again, both the output spectrum and the homogeneity are well simulated using the SISYFOS code, assuming a nonlinear refractive index of $n_2 = 9.3 \times 10^{-23}$ $m^2$/W for Argon and the previously mention value for the FS plate. Only the nonlinear medium dispersion was considered (GVD (FS) = 18.97 $fs^2$/mm and GVD (Ar) = 0.016 $fs^2$/mm), assuming a flat phase from the mirror reflection. For the solid-based broadening, the cell was running in vacuum to avoid any influence of the ambient air. The output pulses were compressed to sub-60 fs, in both cases, using chirped mirrors delivering -3200 $fs^2$ of GDD. The output duration and the corresponding FTL are plotted in Figure 3b).

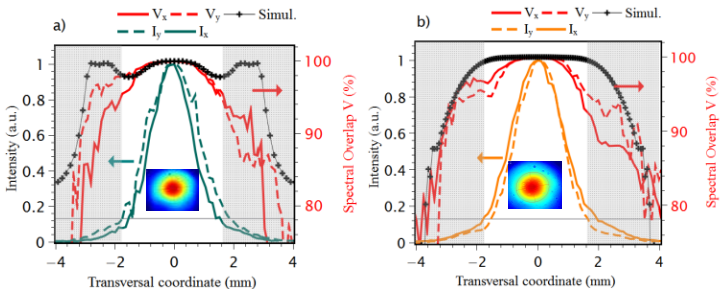

**Figure 4.** Relative intensity (left axis) and spectral overlap V (right axis) across the beam profile along both transverse axes for the vex-cav configuration using 200µJ of input energy in case of: a) 1 mm FS plate as nonlinear medium; b) 7 bar of Argon. Simulation, using SISYFOS code, of the spectral overlap across the radial profile shown in black. Indicative lines showing $1/e^2$ intensity level of the output beam.

In conclusion, we proposed a novel multi-pass configuration for the spectral broadening and compression of ultrashort pulses. In two proof-of-principle experiments, we demonstrate compression, of pulses with 15 µJ and 200 µJ of energy, from 260 fs to sub-60 fs at 15 W average power with an efficiency above 90 %. Excellent spatio-spectral homogeneity of the output beam was measured for both solid- and gas-based broadening. The vex-cav multi-pass configuration extends the onset of SPM in air and the peak power limit for the broadening in bulk material. We foresee that the main implication of the energy scaling concept will be the extension of spectral broadening and compression toward J-level pulse energies, circumventing the current gas ionization and optical damage threshold limitations.

**Funding.** This work is partially supported by the Fundação para a Ciência e a Tecnologia (grant agreement No. PD/BD/135222/2017).

**Acknowledgments.** We would like to thank Gunnar Arisholm for providing his assistance with the simulations and to sincerely acknowledge a few facilitators: A. Borchers, D. Kiesewetter, and A. Puckhaber.

**Disclosure** The authors declare no conflicts of interest.

**Data availability** Data available upon reasonable request from the authors. See Supplement 1 for supporting content.